International Journal of Computer Science, Engineering and Information Technology (IJCSEIT), Vol. 4, No.2, April 2014# A New Multi-Tiered Solid State Disk Using SLC/MLC Combined Flash Memory

Arash Batni[1] and Farshad Safaei[2,1]

[1]Faculty of ECE, ShahidBeheshti University G.C., Evin 1983963113, TEHRAN, IRAN
[2]Institute for Studies in Theoretical Physics and Mathematics (IPM)
School of Computer Science, TEHRAN, IRAN## ABSTRACT

*Storing digital information, ensuring the accuracy, steady and uninterrupted access to the data are considered as fundamental challenges in enterprise-class organizations and companies. In recent years, new types of storage systems such as solid state disks (SSD) have been introduced. Unlike hard disks that have mechanical structure, SSDs are based on flash memory and thus have electronic structure. Generally a SSD consists of a number of flash memory chips, some buffers of the volatile memory type, and an embedded microprocessor, which have been interconnected by a port. This microprocessor run a small file system which called flash translation layer (FTL). This software controls and schedules buffers, data transfers and all flash memory tasks. SSDs have some advantages over hard disks such as high speed, low energy consumption, lower heat and noise, resistance against damage, and smaller size. Besides, some disadvantages such as limited endurance and high price are still challenging. In this study, the effort is to combine two common technologies - SLC and MLC chips - used in the manufacture of SSDs in a single SSD to decrease the side effects of current SSDs. The idea of using multi-layer SSD is regarded as an efficient solution in this field.*## KEYWORDS

*Solid state disk (SSD), Flash Memory, single-level cell (SLC), multi-level (MLC), flash translation layer (FTL), multi tiered cell*

## 1. INTRODUCTION

The high price of solid state disks (SSDs) versus high energy consumption and possibility of wreckage in the traditional hard disks in portable computers or data centers has made the feasibility of reaching a cheap yet reliable storage system too difficult [1-3]. In order to remedy such issues, the need to use a combination of disks or cheap SSDs comparable with traditional hard disks is increasing everyday indata centers and this even holds for general applications.

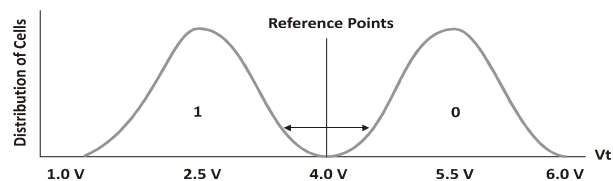

Figure 1. Reference voltage diagram in a single level cell (SLC) [4]

Using flash memory based SSDs instead of hard disks poses several limitations toward the efficiency and reliability criteria by itself. Since flash memory compared to hard disk has an

DOI : 10.5121/ijcseit.2014.420211

intrinsically lower writing speed of data, using flash memories alone in SSDs will degrade the efficiency of the systems solely based on SSDs compared with traditional systems based on hard disk. Hence, various techniques are employed to reduce access latency of flash memories in SSDs in order to improve efficiency of the disks to a level compatible with hard disks. For the time-being, there are two major technologies for the production of flash memories called single-level cell (SLC) and multi-level (MLC) each of which has solved a weakness of SSDs. As shown in Figure 1, since the SLC only keeps one bit per transistor, which means just one threshold voltage is kept for the recognition of "0" and "1", therefore it has a higher reliability.

Whereas in MLC, each transistor represents more than a single bit this is achieved by increasing the number of threshold voltages [5]. Therefore, MLCs have more density and lower prices (at least two times lower) and at the same time show lower reliability and write cycles (see Figure 2).

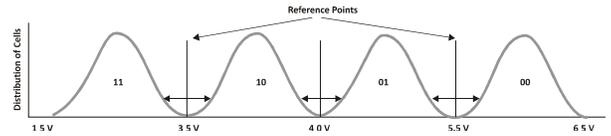

Figure 2. Reference voltage diagram in a multi-level cell (MLC) [4]

The interesting point is that both of the technologies employ similar manufacturing processes and techniques. The differences are in terms of characteristics and efficiency as mentioned in Tables 1 and 2.

Table 1. Structural characteristics of the two technologies used in SSDs [6]

| Characteristic | Single-level cell | Multi-level cell |
|---|---|---|
| High density in the chip level unit |  | ● |
| Lowest cost per bit |  | ● |
| Highest stability | ● |  |
| Lowest temperature in active mode | ● |  |
| Highest writing/erasing speed | ● |  |
| Highest writing/erasing cycle | ● |  |

Table 2. Effective characteristics in the performance of the two technologies used in SSDs [6]

| Characteristic | Single-level cell | Multi-level cell |
|---|---|---|
| Page size | 4 Kilo bytes ||
| Block size | 256 Kilo bytes (256 pages) | 512 Kilo bytes (128 pages) |
| Page reading speed | 45 microseconds | 50 microseconds |
| Writing on page speed | 240 microseconds | 1 millisecond |
| Block deleting speed | 500 microseconds ||
| Maximum writing times on each block | 100 times | 10 thousand times |

SSDs are much more expensive than hard disk in terms of the final cost per gigabyte of storage space. In 2012, the price per gigabyte of storage space (see Figure 3) was $1 versus $0.054 for SSDs and hard disks, respectively [7].

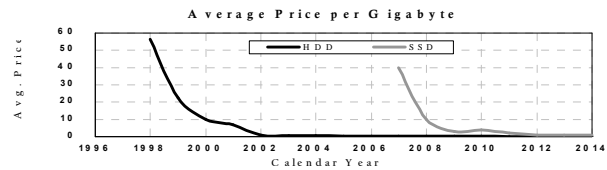

Figure 3. The average price of solid state disk (SSD) and traditional disk per gigabyte [7]

So far, all commercial products in the field of SSDs have been provided merely based on one of the two aforesaid technologies. The idea of using multi-level SSDs is considered as an efficient strategy in the field. Besides preventing the increase of final price and the number of chips in mass volume, due to using multi-level cells several times more than single-level cells, for the storage of vital and useful information (with high number of updates) in single-level cells removes the basic flaws of SSDs versus hard disks.

## 2. RELATED WORK

Now, for the sake of saving cost, energy, and redesign time in computerarchitecture, usually heterogeneous memories such as SRAM, DRAM, and non-volatile memories are replaced by the traditional methods. The main issue with the aforementioned traditional methods is the management of heterogeneous memories. In other words, the issue is about placement or migration of data among different types of memories based on a specific algorithm to access data [8, 9].

Avissar et al. [10], by portioning the data with time compile technique between SRAM, local DRAM, shared DRAM, and ROM showed that this method can significantly improve the running time of programs. Also in 2007, two industry giants, Samsung and Seagate introduced a prototype of combined disk compromised of a traditional hard disk and a SSD on which the specific orders of the operating system were performed thanks to the high speed of reading operation SSDs [11]. Later, other vendors provided industrial instances of this type of disk. In the category of similar technologies we can refer to Microsoft's ReadyBoost™ technology, which employs user's flash memory as a faster unit compared with hard disk to perform the orders of the operating system. Another technology worth to mention is the TurboMemory™ technology of Intel which employs a proprietary 512 megabyte to 1 gigabyte flash memory.

Moreover, in perspective of managing and reducing power consumption, Kim et al. [12], by placing specific information on flash chips instead of the traditional hard disks, and also Bisson [13] and Chen [14], by using USB flash disks as the cache of traditional hard disks, were successful.

In terms of resource management, some work have been reported to reduce special expenses involved in cloud computing and in data centers including Zhang [15, 16] and Akaike et al. [17], who presented a method of data migration in new data centers which employs a set of SSDs as layer zero for storage. However, the idea of using a combination of chips for the first time was presented by Chang and his colleagues [8] which was based on the following reasons:

- Manufacturing affordable SSDs by using MLC and SLC with a ratio larger than the single-surface chip
- Increasing data reliability by migrating vital data to single-surface chips
- Achieving higher writing speed by transferring useful data to the SLCs
- Managing power consumption by transferring data with fewer return times per time unit than the MLC

All these benefits can only be realized by having a very intelligent management unit capable of answering the following questions:

- What is useful (hot) data?
- When should the transfer from the MLC to the SLC be done?
- 

The more parameters flash translation layer of flash memory uses the more precise the final answer will be and thereby the overall efficiency will improve. For example, one of the weaknesses of Zhang's paper [7] is the assumption that useful (hot) data have less size and instead the possibility of updating larger data will be less. This act of relating algorithm data to workload has caused the weakness such that there are many workloads which deal with updating of large data which in case are kept in the surface of multi-level cells, will reduce the lifespan of the disk.

Therefore, in order to improve the efficiency of multi-level disks, the history of last update for each block should be considered; so that if updated, the counter will increase by one unit and in case of reaching a threshold level (which can be different in terms of the workload type) data migration from the MLC to the single-level one will happen or the opposite way. Of course, this management is not only limited to counter, but also the number of classifications and data storage classes can be divided based on parameters of data storage systems (delay, bandwidth, reliability level, availability level and etc) and data can be referred to the intended unit in terms of on demand or automatic detection. The eventual output of this paper is to design a SSD with translation layer unit of flash memory, which in terms of efficiency is capable of competing with traditional SSDs. In other words, the steps of this study includes the following items:

- Achieving an accurate ratio of the number of single-level and MLCs. The aforesaid ratio must be capable of competing with multi-level disks by maintaining the high reliability of single-level disks, and simultaneously providing benefits of the large amount of surface area and competitive final price.
- Designing an efficient layout for MLCs
- Designing flash memory translation layer with a more intelligent information management capability

## 3. MULTI-TIERED SSD

The proposed algorithm called Adaptive Hot Data Migration (AHDM) is an algorithm based on dynamic wear-leveling [18] which adapts the amount of migration threshold in terms of workload type. The primary goal of this method is to combine MLCs and SLCs in order to achieve a longer life span and higher speed with lower price than single-level disks. The flowchart of the proposed algorithm is demonstrated in Figure 4.

The cost of multi-level disks is much less than single-level disks. However, multi-level-disks have higher latency and shorter span compared to their single-level counterparts. By adding one (or a small number of SLCs compared to the number of MLCs) SLC the latency and life span close to the MLCs of the set can be achieved. That means, identifying hot data and referring them to SLC prevents depreciation of MLCs and processes the requests with less latency.

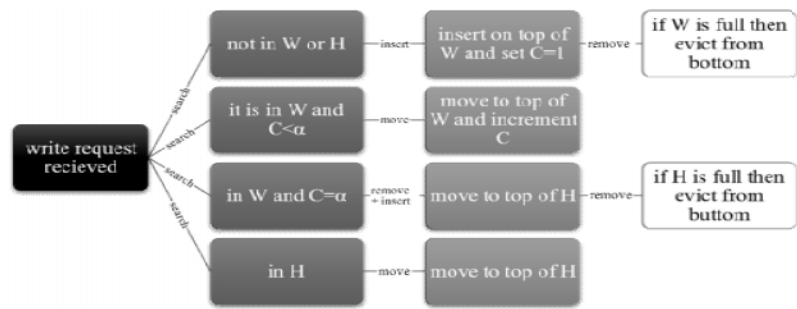

Figure 4.Flowchart of the proposed algorithm

By the arrival of the request, firstly, the type of request must first be examined. If it is of reading type, it will be processed routinely. However, if it is of writing type, it must be compared with the elements in lists W and H. If it cannot be found in any of aforementioned lists, it should be added to the beginning of the list W. If it is in W, we should look at the number of its referrals, if less than the threshold, we just move it to the top of the list and increase the number of its referrals. Nevertheless, if it was equal to the threshold value it should be removed from the list W and be transferred to list H. If the list H is full, we should remove the last element to make space for the new element. Finally, if it is in list H, it is just transferred to the beginning of the line.

## 3.1 AHDM Implementation on Disksim 4.0 Simulator

In the suggested algorithm, two LRU queues are used for the identification of hot data. One for hot data (list H) and the other for warm data (list W). Requests are performed on a single page and pages in list H are stored on SLC. All other pages are stored on MLCs. For addresses in list W, we also keep the number of referrals. If the value of parameter is more than the limit of migration threshold coefficient, the page is transferred to list H and moved from MLC to SLC. If the address of new request is in lists W and H, then the unit will be moved to the top of the list. The use of LRU list causes referral frequency and freshness of data to be involved in deciding whether to move to the SLC or not.

For LRU list, we can use various data structures. Including queues, linked lists, or doubly linked lists along hash table. The cost and complexity of different functions of such data structures have been compared in Tables 3 and 4. As we can see, by utilizing a two-way linked list against the queue, the time complexity will be much improved. However, the required space for the third method is much more than the two previous cases. Therefore, in order to prevent code complexity and to reduce the implementation cost and the required memory, the slight benefit of hash table was ignored so we implement the two-way linked list for storing the data.

Table 3. Time overhead of various data structures[*]

|  | Queue | Doubly Linked List | Linked List + Hash Table |
| --- | --- | --- | --- |
| Insert on the top | O(1) | O(1) | O(1) |
| Evict from the bottom | O(1) | O(1) | O(1) |
| Move to the top | O(n) | O(1) | O(1) |
| Search an element | O(n) | O(n) | O(1) |
| Remove an element | O(n) | O(1) | O(1) |

* in the worst-case *n* is equal to the queue length

International Journal of Computer Science, Engineering and Information Technology (IJCSEIT), Vol. 4, No.2, April 2014

Table 4.Capacityoverhead of different data structures[*]

| | |
|---|---|
| **Queue** | O(n) |
| **Double Linked List** | O(n) |
| **Linked List + Hash Table** | O(number of SSD pages + n) |

* in the worst-case *n* is equal to the queue length

## 4. EXPERIMENTAL RESULTS

### 4.1 DISTRIBUTION OF THE REQUESTS

The simulation results of the proposed algorithm on four workloads have been plotted on Figure 5. The figure shows that the data migration threshold works well with different working conditions and transfers hot data to the SLC. Thereby, the number of writings on MLCs, for tasks that a large fraction of their requests are writing, reduces in half, and their lifespan will increase by the same amount.

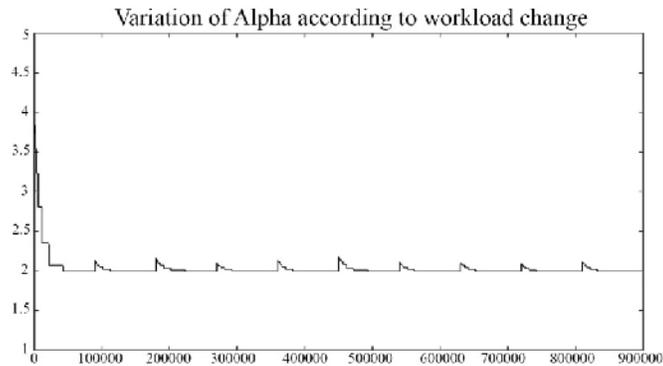

Figure 5. The variation of data migration threshold coefficient for IOzone workload

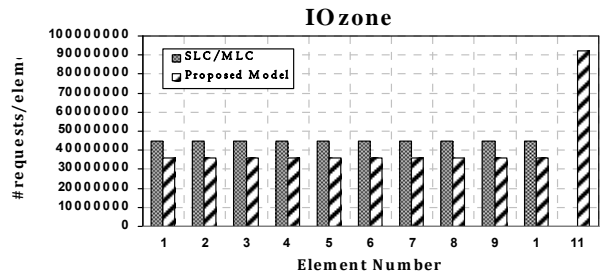

Figure 6. Distribution of the replied requests in IOzone workload

As shown in Figure 6, due to the fact that the majority of requests in the workload are write applications, 20.49% of the hot requested data is identified and migrated.

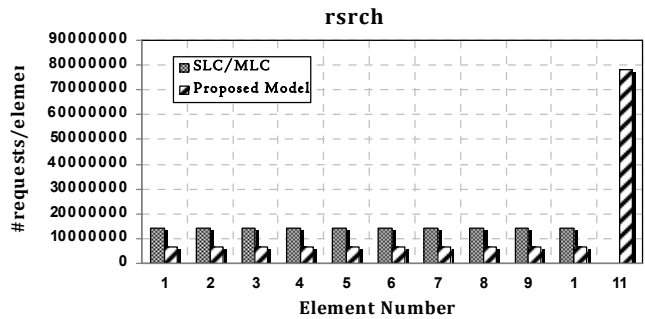

Figure 7. Distribution of the replied requests in rsrch workload

Most of the requests of rsrch workload are like IOzone of writing type; but because the writing requests to the similar addresses in rsrch workload are much more than IOzone workload and as seen in Figure 7, 54.79% of the requested data are identified and migrated.

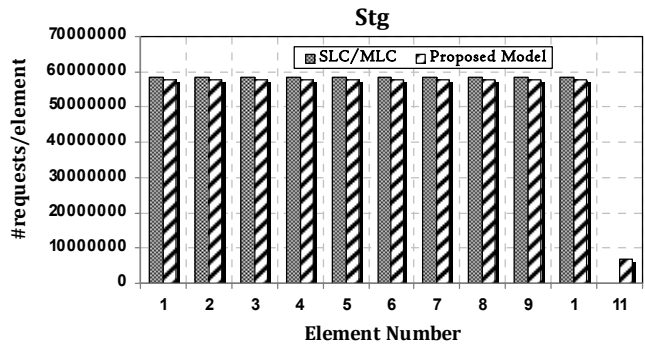

Figure 8. Distribution of the replied requests in stg workload

In the two other workloads, i.e. stg and web, since most of the requests are of reading type, which means we should not expect a similar performance, that is why in stg workload, only 1.19% and in web, 0.07% of data is transferred to the single-level cell (see Figure 8).

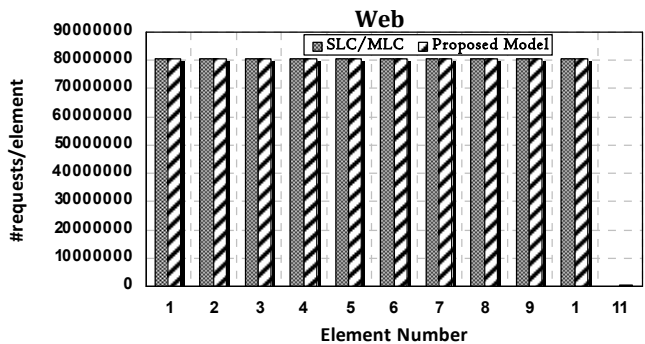

Figure 9. Distribution of the replied requests in web workload

If we put the results of the web workload aside (since writing constitutes more than 99% of requests) on average 25.49% of the data is migrated in the three workloads (see Figure 9).

## 4.2 THE AVERAGE NUMBER OF WRITINGS ON THE BLOCK

The results of this test are the nearest way to show improvement of chip lifespan. In order to be able to refer to the simulation results, on average, we should write more than 30 times on each chip block. This is regardless of the web workload results.

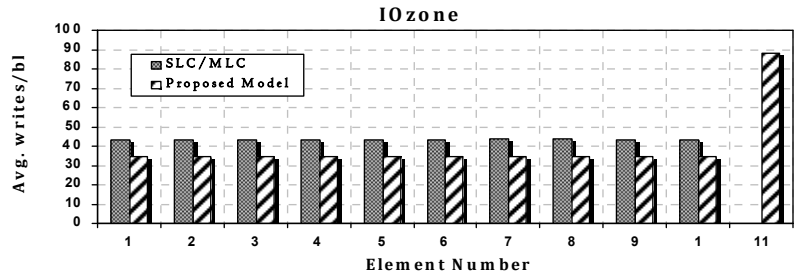

Figure 10. The average number of writing on block in IOzone workload

According to Figure 10, the average level of writings on blocks of MLCs have decreased from 43.49 times to 34.55 times which shows a 20.56% increase in the lifespan of MLCs. Further, each block of single-level chip, on average, is written on 88.35 times that is 2.5 times higher than of MLCs.

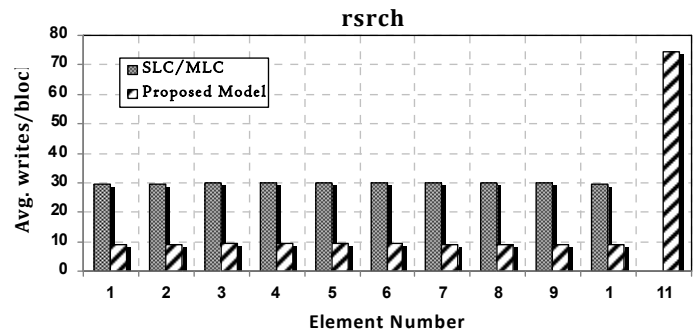

Figure 11. The average number of writings on a block in rsrch workload

Since the number of writings in rsrch workload in similar addresses is higher, the algorithm identifies more hot data and thereby the average number of writings in Figure 11 shows larger decrease. The average number of writings on blocks of the MLCs is decreased from 29.73 to 9.19 which shows a 69.09% increase in the lifespan of MLCs. Moreover, on average, each block of SLC is written on 74.33 times which reveals an 8 times higher number of writings than the MLC which is very satisfactory.

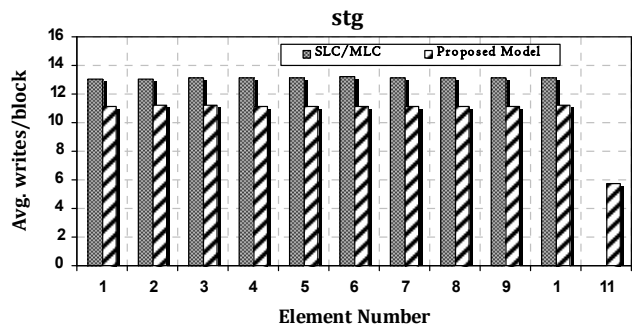

Figure 12. The average number of writings on blocks in stg workload

Figure 12 illustrates the average number of writings on blocks with stg workload fortified with the proposed algorithm. The average number of writings is decreased from 13.1 to 11.15. The reason behind the 14.88% decrease lies in the fact that the most of requests of rsrch workload are of reading type and the purpose of choosing such a workload was to show the readiness of the proposed algorithm to reply for all requested ranges. Simulation results show that the overall average increases by 34.84% over the lifetime of the chips are proper. In the case of general workloads, such results are considered acceptable.

### 4.3 THE ACCESS TIME

The ordinary time of access to the data residing on slash-based SSD is about 25 to 100 microseconds. While the speed of the HDDs ranges from 5,000 to 10,000 microseconds. Therefore, the SSDs operate approximately 100 times faster than HDDs. HDDs typically transfer data with the speed of 80 to 120 megabytes per second. Whereas, the data transfer rate of SSDs is 170 to 250 megabytes per second. With the lack of moving parts, SSDs' access times are much better than the traditional hard disks, although the difference between single surface and multi surface disk is very high, the proposed algorithm can fill this gap.

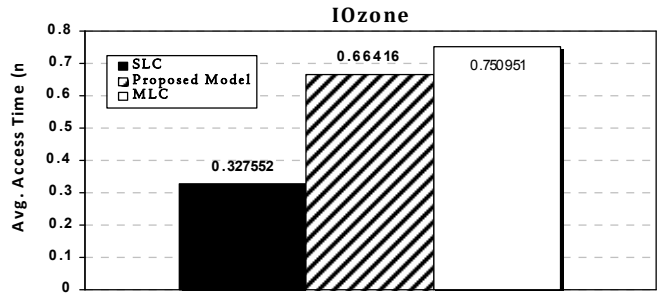

Figure 13. The average data access time in IOzone workload

As shown in Figure 13, the access time in the proposed algorithm demonstrates a 11.56% decrease compared to the pure MLCs. However, this decrease is not big enough to be considered close to the SLC but still with regard to the workload type, the results seem acceptable.

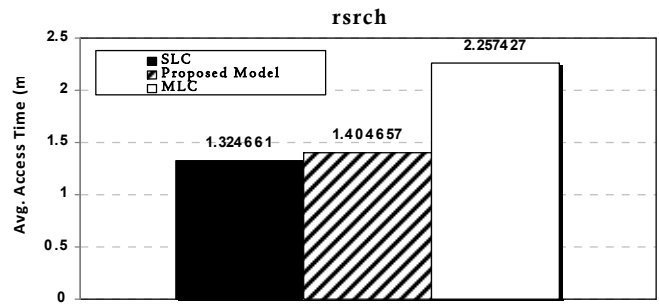

Figure 14. The average data access time in rsrch workload

The results of data access time for rsrch workload just like previous tests are better than the other workloads so that we have a 37.78% decrease of the access time; thereby, it can be said that the results are compatible with the SLC (see Figure 14).

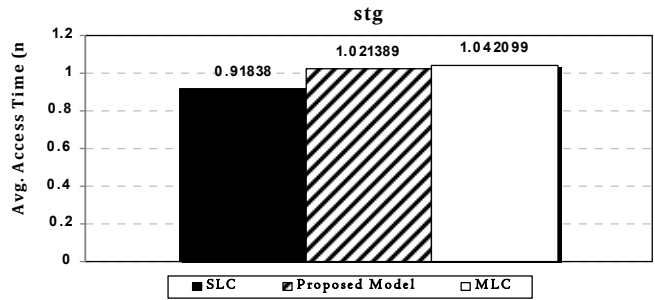

Figure 15. The average data access in stg workload

As depicted by Figures 15 and 16, in the case of stg and web workloads, due to the content of the requests, there is not much discrepancy between the single-level and the MLCs. Hence, the proposed algorithm does not really seem so efficient, such that in stg workload, there was a 1.98% decrease and in regard to web workload there was almost no change compared to its multi-level counterpart.

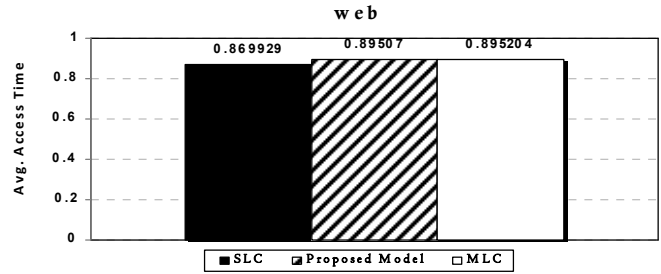

Figure 16. The average data access time in web workload

Overall, the overview of the conducted tests shows that the average access time is reduced by 17.11%.

## 5. DISCUSSIONS

### 5.1 COMPARING THE PRICES

One of the purposes of this study is to provide a low cost solution for improving the performance of MLCs. For the fare comparison between the proposed disk price with SSDs based on SLC and MLC we prepared Table 5 based on the prices reported in [19]. Moreover, for the better understanding Figure 17 compares the total cost of these three techniques. As cleared by this figure, the proposed model has only 25% increase in price in comparison with MLC based disks.

Table 5. Comparing the final price of tested disks[*]

| The proposed model | MLC | SLC | Chip type |
|---|---|---|---|
| – | $0.9 | $3.00 | **Chip price** |
| $12.00 | $9.00 | $30 | **Total price** |

*According to the price list of OCZ company chips in November 2013 [19]

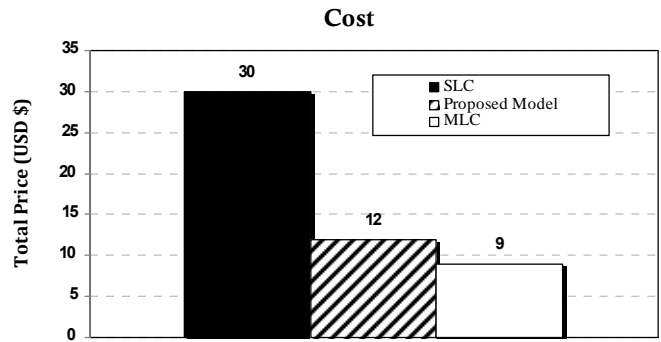

Figure 17. Comparing the price of SLC- and MLC- based SSD with proposed model

## 5.2 THE AVERAGE ACCESS TIME

Figure 18 plots the average access times of the four workloads utilized in the simulation experiments. It is apparent from the figure, in the proposed model by spending a low cost (>60% lower than the cost of the pure SLC model); the results obtained by the simulation are close to the single-level disk (only 14% slower).

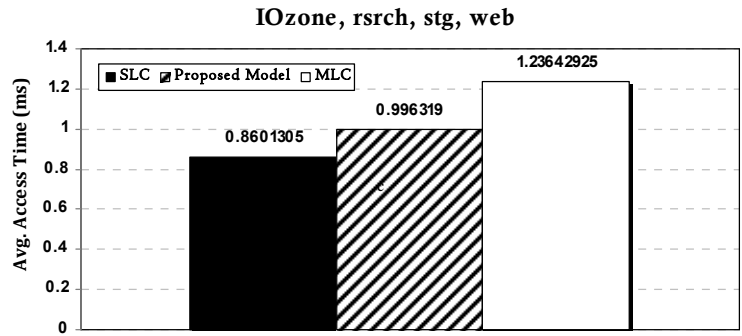

Figure 18. The average access time in the four related workloads

## 5.3 THE LIFETIME OF THE DISK

The decrease in the number of writings on each chip blocks will increase its lifetime. Therefore, based on the average workloads, the lifetime of the proposed method shows a 35% increase over the multi-level model.

## 5.4 THE RATIO OF CHIPS IN THE PROPOSED MULTI-LEVEL MODEL

Another objective of this study is to determine an accurate ratio of the single-level and the MLCs in the proposed multi-layer model. As can be seen in Figures 13 ~ 18 the ratio of the single-level and the MLCs, nevertheless of the voluminous nature of the workloads the lifespan of the blocks in the SLC is about 2.5 times of MLCs. Since the average lifespan of the SLCs is 10 times higher than the MLCs, the ratio of utilizing chips in the multi-layer model is increasable to 1 by 40.

Another limitative factor is the saturation of SLC, which as a result of overly writings in the ratio of 1 to 40, the task failed to finish. Finally, the extensive simulation results of various workloads with respect to the maximum lifetime and the lack of saturation of SLC, recommends the usage ratio of 1 to 10 (≈ 10%) of the SLC.

## 6. CONCLUSIONS

The main objective of this paper was to provide a new approach to the spectrum of current technologies utilized for manufacturing flash chips by bringing positive features along the negative points together. Positive characteristics include low cost in the multi-level cells (MLCs) and low access time and long lifetime in the single-level cells (SLCs). The performance of the proposed method has extensively evaluated, and the simulation experiments confirm that the model exhibits a good degree of accuracy under different workloads and it is valid under various network conditions.

## REFERENCES

[1]  Samsung Electronics Company. K9GAG08U0M 4Gb * 8 Bit NAND Flash Memory Data Sheet (Preliminary).
[2]  Samsung Electronics Company. K9NBG08U5M 4Gb * 8 Bit NAND Flash Memory Data Sheet.
[3]  Samsung Electronics Company. "NAND Flashbased Solid State Disk Data Sheet".
[4]  "SLC vs. MLC: An Analysis of Flash Memory" white paper, Super Talent Technolog.
[5]  F.Roohparvar, "Single level cell programming in a multipl level cell non-volatile memory device," *InUnited States Patent*, No. 7,366,013.
[6]  SoojunIm, Dongkun Shin, "Storage Architecture and Software Support for SLC/MLC Combined Flash Memory", *In SAC '09: Proceedings of the 2009 ACM symposium on Applied Computing,* pp 1664-1669, 2009.
[7]  Data source: Mkomo.com, Gartner, and Pingdom (December 2011).
[8]  L. Chang. Hybrid, "solid-state disks: combining heterogeneous NAND flash in large SSDs," *Design Automation Conference*, *ASPDAC 2008. Asia and South Pacific*, pp 428–433, 2008.
[9]  S.Hong, D.Shin "NAND Flash-based Disk Cache Using SLC/MLC Combined Flash Memory." *SNAPI '10 Proceedings of the 2010 International Workshop on Storage Network Architecture and Parallel I/Os*, pp 21-30, 2010.
[10] OREN AVISSAR et al. "A An Optimal Memory Allocation Scheme for Scratch-Pad-Based Embedded Systems," *ACM Transactions on Embedded Computing Systems*, Vol. 1, No. 1, pp 6–26, November 2002.
[11] Perenson, Melissa. "Tested: New Hybrid Hard Drives from Samsung and seagate",http://www.pcworld.com/article/138102/article.html.
[12] Young-Jin Kim et al. "Energy-Efficient File Placement Techniques for Heterogeneous Mobile Storage Systems," *EMSOFT'06*, October 2006, Seoul, Korea.
[13] Timothy Bisson Scott A. Brandt, "Reducing Energy Consumption using a Non-Volatile storage cache," *IWSSPS held in conjunction with the IEEE Real-Time and Embedded Systems and Applications Symposium (RTAS),* March 2005.
[14] Feng Chen et al. "SmartSaver: Turning Flash Drive into a Disk Energy Saver for Mobile Computers," *ISLPED'06, October 2006*, Tegernsee, Germany.
[15] Gong Zhang et al. "Automated Lookahead Data Migration in SSD-enabled Multi-tiered Storage Systems" In Mass Storage Systems and Technologies (MSST), 2010 IEEE 26th Symposium on, May 2010.
[16] Gong Zhang et al. "Adaptive Data Migration in Multi-tiered Storage Based Cloud Environment," *IEEE 3rd International Conference on Cloud Computing*, July 2010.
[17] HirotoshiAkaike et al. "Performance Evaluation of Energy-efficient High-speed Tiered-storage System," *Industrial Informatics (INDIN), 2010 8th IEEE International Conference*, July 2010.
[18] Joosung Yun et al. "Bloom filter-based dynamic wear leveling for phase-change RAM." Design, Automation & Test in Europe Conference & Exhibition (DATE), pp 1513 – 1518, 2012,
[19] "OCZ SLC/MLC chip price" available on: *http://ocz.com/*

## Authors

**ArashBatni**He received the B.Sc. degree from Shahed University, Tehran, Iran, in 2010, and the M.Sc. degree from ShahidBeheshti University (SBU), Tehran, Iran, in 2013, both in computer engineering. His current research interests include Solid State Disks, Storage Systems, Distributed Parallel File Systems, High Performance Computing and GPU Parallel Programming.

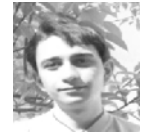

**FarshadSafaei** He received the B.Sc., M.Sc., and Ph.D. degrees in Computer Engineering from Iran University of Science and Technology (IUST) in 1994, 1997 and 2007, respectively. He is currently an assistant professor in the Department of Electrical and Computer Engineering, ShahidBeheshti University, Tehran, Iran. His research interests are performance modelling/evaluation, Interconnection networks, social and complex networks, and high performance computer architecture.

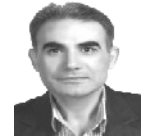